\documentclass{jfm}
\usepackage{graphicx}
\usepackage{epstopdf}
\usepackage{float}
\usepackage{morefloats}
\usepackage{placeins}
\usepackage{xfrac}
\usepackage{pdflscape}
\usepackage{hhline}
\usepackage[table]{xcolor}
\usepackage{rotating}
\definecolor{light}{cmyk}{0,0,0,0.10}
\usepackage{tabularx}
\newcolumntype{Y}{>{\centering\arraybackslash}X}

\begin{document}

{ 
\newtheorem{lemma}{Lemma}
\newtheorem{corollary}{Corollary}

\shorttitle{Prandtl effects on turbulent sheared thermal convection} 
\shortauthor{A. Blass \etal} 

\title{The effect of Prandtl number on turbulent sheared thermal convection}

\author
{
	Alexander Blass\aff{1}
	\corresp{\email{a.blass@utwente.nl}},
    Pier Tabak\aff{1},
	Roberto Verzicco\aff{2,1,3},\\
	Richard J.A.M. Stevens\aff{1},
	\and 
	Detlef Lohse\aff{1,4}
	\corresp{\email{d.lohse@utwente.nl}}
}

\affiliation
{
	\aff{1}
	Physics of Fluids Group, Max Planck Center for Complex Fluid Dynamics, J. M. Burgers Center for Fluid Dynamics and MESA+ Research Institute, Department of Science and Technology, University of Twente, P.O. Box 217, 7500 AE Enschede, The Netherlands
	\aff{2}
	Dipartimento di Ingegneria Industriale, University of Rome "Tor Vergata". Via del Politecnico 1, Roma 00133, Italy
	\aff{3}
	Gran Sasso Science Institute - Viale F. Crispi, 7 67100 L'Aquila, Italy.
	\aff{4}
	Max Planck Institute for Dynamics and Self-Organization, Am Fassberg 17, 37077 G\"ottingen, Germany
}

\maketitle
}

\begin{abstract}
In turbulent wall sheared thermal convection, there are three different flow regimes, depending on the relative relevance of thermal forcing and wall shear. In this paper we report the results of direct numerical simulations of such sheared Rayleigh-B\'enard convection, at fixed Rayleigh number $Ra=10^6$, varying the wall Reynolds number in the range $0 \leq Re_w \leq 4000$ and Prandtl number $0.22 \leq Pr \leq 4.6$, extending our prior work by \cite{bla20}, where $Pr$ was kept constant at unity and the thermal forcing ($Ra$) varied. We cover a wide span of bulk Richardson numbers $0.014 \leq Ri \leq 100$ and show that the Prandtl number strongly influences the morphology and dynamics of the flow structures. In particular, at fixed $Ra$ and $Re_w$, a high Prandtl number causes stronger momentum transport from the walls and therefore yields a greater impact of the wall shear on the flow structures, resulting in an increased effect of $Re_w$ on the Nusselt number. Furthermore, we analyse the thermal and kinetic boundary layer thicknesses and relate their behaviour to the resulting flow regimes. For the largest shear rates and $Pr$ numbers, we observe the emergence of a Prandtl- von Karman log-layer, signalling the onset of turbulent dynamics in the boundary layer. Finally, our results allow to extend the Grossmann-Lohse theory for heat transport in Rayleigh-B\'enard convection to the sheared case, universally describing $Nu(Ra,Pr,Re_w)$.
\end{abstract}

\section{Introduction}
Buoyancy and shear are crucial processes in fluid dynamics and key for many flow related processes in nature and technology. A paradigmatic example of buoyancy driven flow is Rayleigh-B\'enard (RB) convection, a system where the fluid is heated from below and cooled from above \citep{ahl09,loh10,chi12,xia13}. The flow is controlled by the Rayleigh number $Ra=\beta gH^3\Delta /(\kappa\nu)$, which quantifies the the non-dimensional temperature difference between the two horizontal plates. Here, $H$ is their distance, $\beta$ the thermal expansion coefficient of the fluid, $g$ the gravitational acceleration, $\Delta$ the temperature difference across the fluid layer, $\kappa$ and $\nu$ the thermal diffusivity and kinematic viscosity, respectively. Furthermore, the Prandtl number is defined as $Pr=\nu/\kappa$, which is the ratio between momentum and thermal diffusivities. An important output of the flow is the heat transport between the plates, which can be non-dimensionally quantified by the Nusselt number $Nu= QH/(\kappa \Delta)$, with $Q=\overline{w' T'} - \kappa \partial T /\partial z$ the mean vertical heat flux ($w'$ and $T'$ are vertical velocity and temperature fluctuations, z is the vertical direction). 

On the other hand, for flows driven by wall shear stress, a commonly used model problem is the Couette flow \citep{thu00,bar05,tuc11}. We adopt a geometry in which the bottom and top walls slide in opposite directions with a wall-tangential velocity $u_w$ and the forcing can be expressed non-dimensionally by the wall Reynolds number $Re_w=H u_w/\nu$. The relevant flow output is now the wall friction, quantified by the friction coefficient $C_f = 2\tau_w/(\rho u_w^2)$, with $\rho$ the fluid density and $\tau_w$ the surface- and time-averaged wall shear stress. Turbulent Couette flow is dominated by large-scale streaks \citep{lee91,tsu06,kit08,pir11,pir14,orl15,cha17}. These remain correlated in the streamwise direction for a length up to about $160$ times the distance between the plates \citep{lee18}. 

Combining both, buoyancy and wall shear forcings, yields a complex system that is relevant in many applications, especially for atmospheric and oceanic flows \citep{dea72,moe84,kha98}. Also in sheared thermal convection large-scale structures emerge, as experiments have shown \citep{ing66,sol90}. Investigations on channel flows with unstable stratification \citep{fuk85} revealed that temperature fluctuations in the bulk decrease while velocity fluctuations close to the wall increase for stronger unstable stratification. 

Numerical simulations of wall sheared convection \citep{hat86,dom88} have revealed that adding shear to buoyancy increases the heat transport for low $Ra$, but causes also the large-scale structures to weaken thus decreasing the heat transport for $Ra \gtrsim 150.000$. Similar phenomena have been observed in Poiseuille-RB, where the wall parallel mean flow is driven by a pressure gradient rather than the wall shear: in this case the $Nu$ decrease was attributed to the disturbance of the longitudinal wind on the thermal plumes \citep{sca14,sca15,pir17}. This plume-sweeping mechanism, causing a Nusselt number drop, was also observed in \cite{bla20}, who report very long, thin streaks, similar to those of the atmospheric boundary layer where these convection rolls are called cloud streets \citep{etl93,kim03,jay18}.

In both flows, Couette-RB and Poiseuille-RB, the ratio between buoyancy and mechanical forcings can be best quantified by the bulk Richardson number 
\begin{equation}
Ri=\frac{Ra}{Re^2_wPr},
\end{equation}
which is a combination of the flow governing parameters $Ra$, $Re_w$ and $Pr$. In the Couette-RB flow of \cite{bla20}, $Ri$ was in fact used to distinguish between three different flow regimes, namely thermal buoyancy dominated, transitional, and shear dominated, similarly to the case of stably stratified wall turbulence, where \cite{zon18} distinguish between the buoyancy dominated, buoyancy affected and turbulence dominated regimes. 

Indeed, sheared stably or unstably stratified flows are present in many different situations involving both liquids and gases. Therefore the fluid properties, as reflected in the Prandtl number, play a major role \citep{cho18}. In the atmosphere it results $Pr=\mathcal{O}(1)$ while in ocean dynamics $Pr=\mathcal{O}(10)$. However, a much larger $Pr$ variation is found in industrial applications. E.g. $Pr\approx \mathcal{O}(10^{-3})$ for liquid metals \citep{tei17}, which are for example in use for cooling applications in nuclear reactors \citep{usa99} or $Pr\approx \mathcal{O}(10^{3})$ for molten salts or silicone oils \citep{vig15} for high-performance heat exchangers.  

\begin{figure}
	\centering
	\includegraphics[width=\textwidth]{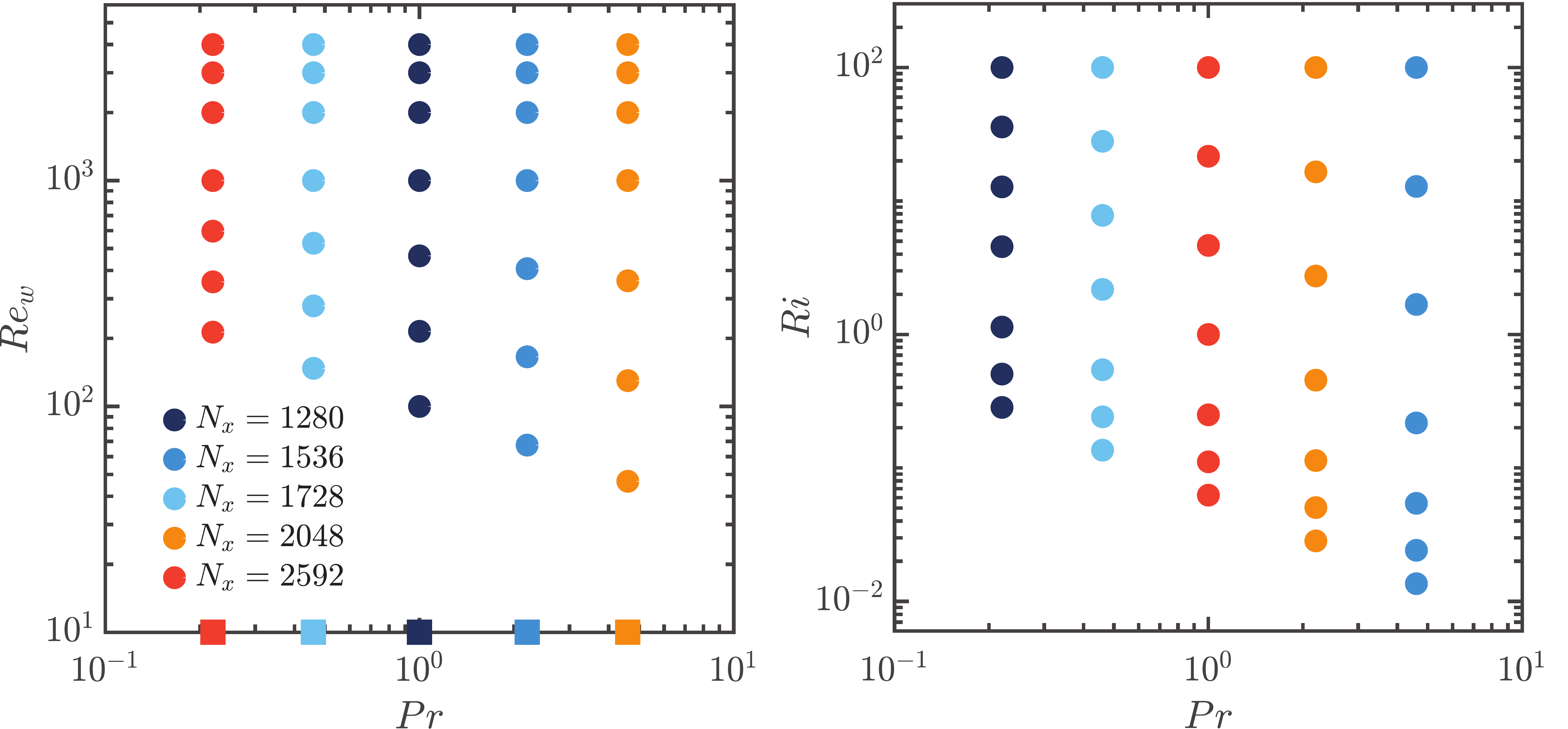}%
	\caption{\label{fig:phase} Phase diagram of simulation runs. We show two panels to better illustrate our choice of simulation input parameters, which were determined based on $Re_w$ (left panel) and $Ri$ (right panel). $Re_w=2000,3000,4000$ were chosen to be consistent with \cite{bla20} and to cover the shear dominated regime. The squared symbols show the datapoints for $Re_w =0$ for completeness and independently of the y-axis, since they cannot be directly included in the logarithmic scale. To have a sufficient amount of data in the thermal buoyancy dominated regime, we picked $Ri=100$ as most thermal dominated case and then logarithmically spaced three more datapoints.}
\end{figure}

Despite this staggering range of Prandtl numbers encountered in real applications, the vast majority of studies on sheared, thermally stratified flows have been performed only at $Pr=\mathcal{O}(1)$. To overcome this limitation, in this paper we extend the work of \cite{bla20} for $Pr=1$ by analysing the parameter space $0\leq Re_w \leq 4000$ and $0.22 \leq Pr \leq 4.6$ while keeping the Rayleigh number constant at $Ra=10^6$ (see figure \ref{fig:phase} for the complete set of simulations).

The present study can be considered similar and complementary to that of \cite{zho17} who carried out numerical simulations with a large $Pr$ variation for a {\it stably} stratified Couette flow.

The manuscript is divided in the following manner. Section \ref{sec:numerical} briefly reports the numerical method. Section \ref{sec:results} focusses on the global transport properties and section \ref{sec:BL} on the boundary layers. The paper ends with conclusions (section \ref{sec:conclusion}).

\begin{table} 
	\vspace{-0.3cm}	
	\begin{center}
		\begin{tabularx}{\textwidth}{c@{}Y@{}Y@{}p{0.3cm}Y@{}Y@{}Y@{}p{0.3cm}Y@{}Y@{}Y@{}YY} 
			$Ra$ & $Pr$ & $Re_w$ & $Ri$ & & $N_x$ & $N_y$ & $N_z$ &  & $Re_{\tau}$ & $L_{MO}/H$ & $Nu$ & $C_f/10^{-3}$ \\[1pt]
			
			\hline \hline \\[-5.5pt]
			
			$ 1.0\times10^{6} $ & 0.22 &     0 & $\infty$ & & 2592 & 2048 & 256 & &  -- &     0 & 7.37 & $\infty$ \\
			$ 1.0\times10^{6} $ & 0.22 &   213 &    100.0 & & 2592 & 2048 & 256 & & 57.90 & 0.006 & 7.33 &    147.5 \\
			$ 1.0\times10^{6} $ & 0.22 &   357 &    35.86 & & 2592 & 2048 & 256 & & 74.80 & 0.013 & 7.24 &    88.28 \\
			$ 1.0\times10^{6} $ & 0.22 &   597 &    12.76 & & 2592 & 2048 & 256 & & 99.81 & 0.031 & 6.98 &    55.90 \\
			$ 1.0\times10^{6} $ & 0.22 & 1000 &    4.546 & & 2592 & 2048 & 256 & & 133.0 & 0.080 & 6.44 &    35.37 \\
			$ 1.0\times10^{6} $ & 0.22 & 2000 &    1.137 & & 2592 & 2048 & 256 & & 197.0 & 0.286 & 5.89 &    19.41 \\
			$ 1.0\times10^{6} $ & 0.22 & 3000 &    0.505 & & 2592 & 2048 & 256 & & 246.7 & 0.538 & 6.14 &    13.53 \\
			$ 1.0\times10^{6} $ & 0.22 & 4000 &    0.284 & & 2592 & 2048 & 256 & & 291.5 & 0.884 & 6.17 &    10.62 \\ [1pt]
			
			\hline \\[-6.5pt]
			
			$ 1.0\times10^{6} $ & 0.46 &     0 & $\infty$ & & 1728 & 1458 & 192 & &  -- &     0 &  7.92 & $\infty$ \\
			$ 1.0\times10^{6} $ & 0.46 &   147 &    100.0 & & 1728 & 1458 & 192 & & 43.75 & 0.005 &  7.82 &    176.1 \\
			$ 1.0\times10^{6} $ & 0.46 &   279 &    27.99 & & 1728 & 1458 & 192 & & 60.63 & 0.014 &  7.58 &    94.69 \\
			$ 1.0\times10^{6} $ & 0.46 &   528 &    7.803 & & 1728 & 1458 & 192 & & 85.08 & 0.041 &  6.98 &    51.97 \\
			$ 1.0\times10^{6} $ & 0.46 & 1000 &    2.175 & & 1728 & 1458 & 192 & & 120.2 & 0.128 &  6.26 &    28.91 \\
			$ 1.0\times10^{6} $ & 0.46 & 2000 &    0.544 & & 1728 & 1458 & 192 & & 175.0 & 0.414 &  5.96 &    15.33 \\
			$ 1.0\times10^{6} $ & 0.46 & 3000 &    0.241 & & 1728 & 1458 & 192 & & 217.8 & 0.787 &  6.04 &    10.54 \\
			$ 1.0\times10^{6} $ & 0.46 & 4000 &    0.136 & & 1728 & 1458 & 192 & & 260.7 & 1.287 &  6.33 &    8.493 \\ [1pt]
			
			\hline \\[-6.5pt]
			
			$ 1.0\times10^{6} $ & 1    &       0 & $\infty$ & & 1280 & 1024 & 256 & &  -- &     0 &  8.34 & $\infty$ \\
			$ 1.0\times10^{6} $ & 1    &   100 &    100.0 & & 1280 & 1024 & 128 & & 31.85 & 0.004 &  8.20 &    202.9 \\
			$ 1.0\times10^{6} $ & 1    &   215 &    21.63 & & 1280 & 1024 & 128 & & 47.31 & 0.014 &  7.82 &    96.86 \\
			$ 1.0\times10^{6} $ & 1    &   464 &    4.645 & & 1280 & 1024 & 128 & & 72.95 & 0.056 &  6.95 &    49.44 \\
			$ 1.0\times10^{6} $ & 1    & 1000 &    1.000 & & 1280 & 1024 & 128 & & 113.5 & 0.223 &  6.56 &    25.75 \\
			$ 1.0\times10^{6} $ & 1    & 2000 &    0.250 & & 1280 & 1024 & 256 & & 161.7 & 0.645 &  6.56 &    13.07 \\
			$ 1.0\times10^{6} $ & 1    & 3000 &    0.111 & & 1280 & 1024 & 256 & & 203.0 & 1.218 &  6.87 &    9.158 \\
			$ 1.0\times10^{6} $ & 1    & 4000 &    0.063 & & 1280 & 1024 & 256 & & 251.7 & 2.022 &  7.89 &    7.922 \\ [1pt]
			
			\hline \\[-6.5pt]
			
			$ 1.0\times10^{6} $ & 2.2  &       0 & $\infty$ & & 1536 & 1296 & 162 & &   -- &     0 &   8.50 & $\infty$ \\
			$ 1.0\times10^{6} $ & 2.2  &     67 &    100.0 & & 1536 & 1296 & 162 & & 22.88 & 0.003 &  8.38 &    230.3 \\
			$ 1.0\times10^{6} $ & 2.2  &   166 &    16.52 & & 1536 & 1296 & 162 & & 37.02 & 0.015 &   7.68 &    99.65 \\
			$ 1.0\times10^{6} $ & 2.2  &   407 &    2.741 & & 1536 & 1296 & 162 & & 63.08 & 0.081 &   6.82 &    47.99 \\
			$ 1.0\times10^{6} $ & 2.2  & 1000 &    0.455 & & 1536 & 1296 & 162 & & 100.4 & 0.336 &   6.62 &    20.18 \\
			$ 1.0\times10^{6} $ & 2.2  & 2000 &    0.114 & & 1536 & 1296 & 162 & & 144.2 & 0.936 &   7.04 &    10.39 \\
			$ 1.0\times10^{6} $ & 2.2  & 3000 &    0.050 & & 1536 & 1296 & 162 & & 194.1 & 1.845 &   8.72 &    8.373 \\
			$ 1.0\times10^{6} $ & 2.2  & 4000 &    0.028 & & 1536 & 1296 & 162 & & 246.1 & 3.052 & 10.75 &    7.573 \\ [1pt]
			
			\hline \\[-6.5pt]
			
			$ 1.0\times10^{6} $ & 4.6  &       0 & $\infty$ & & 2048 & 1536 & 192 & &   -- &     0 &   8.51 & $\infty$ \\
			$ 1.0\times10^{6} $ & 4.6  &     47 &    100.0 & & 2048 & 1536 & 192 & & 16.68 & 0.003 &   8.31 &    255.9 \\
			$ 1.0\times10^{6} $ & 4.6  &   130 &    12.85 & & 2048 & 1536 & 192 & & 29.59 & 0.016 &   7.51 &    103.5 \\
			$ 1.0\times10^{6} $ & 4.6  &   360 &    1.678 & & 2048 & 1536 & 192 & & 53.01 & 0.101 &   6.77 &    43.38 \\
			$ 1.0\times10^{6} $ & 4.6  & 1000 &    0.217 & & 2048 & 1536 & 192 & & 87.65 & 0.459 &   6.75 &    15.36 \\
			$ 1.0\times10^{6} $ & 4.6  & 2000 &    0.054 & & 2048 & 1536 & 192 & & 137.1 & 1.382 &   8.58 &    9.397 \\
			$ 1.0\times10^{6} $ & 4.6  & 3000 &    0.024 & & 2048 & 1536 & 192 & & 189.0 & 2.685 & 11.56 &    7.936 \\
			$ 1.0\times10^{6} $ & 4.6  & 4000 &    0.014 & & 2048 & 1536 & 192 & & 240.4 & 4.441 & 14.39 &    7.225 \\ [1pt]
			
			\hline \hline \\[-6.5pt]
			
		\end{tabularx}
		\caption{Main simulations considered in this work. The columns from left to right indicate the input and output parameters and the resolution in streamwise, spanwise, and wall-normal direction $(N_x,N_y,N_z)$. The simulations for $0\leq Re_w \leq 1000$ were chosen to allow the first nonzero $Re_w$ at $Ri=100$. The other two $Re_w<1000$ simulations for each $Pr$ respectively were logarithmically evenly spaced in $Re_w$. Data of \cite{bla20} have been used for $Pr=1;Re=0,2000,3000,4000$. The data of the Monin-Obukhov length was added for consistency with \cite{bla20}, although not specifically discussed in this manuscript.}
		{\label{tab:gridcases}}
	\end{center}
\end{table}

\section{Numerical method}
\label{sec:numerical}
The three-dimensional incompressible Navier-Stokes equations with the Boussinesq approximation are integrated numerically. Once non-dimensionalised, the equations read:

\begin{equation} 
\frac{\partial \boldsymbol{u}}{\partial t} + \boldsymbol{u} \bcdot \bnabla \boldsymbol{u} =-\bnabla P + \left(\frac{Pr}{Ra} \right)^{1/2} \nabla^2\boldsymbol{u}+\theta \hat{z}, \mbox{~~~~} \bnabla \bcdot \boldsymbol{u} =0,
\label{eqn:Navier}
\end{equation}
\begin{equation} 
\frac{\partial \theta}{\partial t} + \boldsymbol{u} \bcdot \bnabla \theta = \frac{1}{(Pr Ra)^{1/2}} \nabla ^2 \theta,
\label{eqn:temp} \\
\end{equation}
with $\boldsymbol{u}$ and $\theta$ the velocity, normalized by $\sqrt{g \beta \Delta H}$, and temperature, normalized by $\Delta$, respectively. $t$ is the time normalized by $\sqrt{H/(g \beta \Delta)}$ and $P$ the pressure in units of $g \beta \Delta /H$. 

Equations (\ref{eqn:Navier}) and (\ref{eqn:temp}) are solved using the AFiD GPU package \citep{zhu18b} which bases on a second-order finite-difference scheme \citep{poe15c}. The code has been validated and verified several times \citep{ver96,ver97,ver03,ste10,ste11,ost14d,koo18}. We use a uniform discretization in horizontal, periodic directions and a non-uniform mesh, with an error function-like node distribution in the wall-normal direction. 

\begin{figure}
	\centering 
	\rotatebox{90}{\includegraphics[width=0.9\textheight]{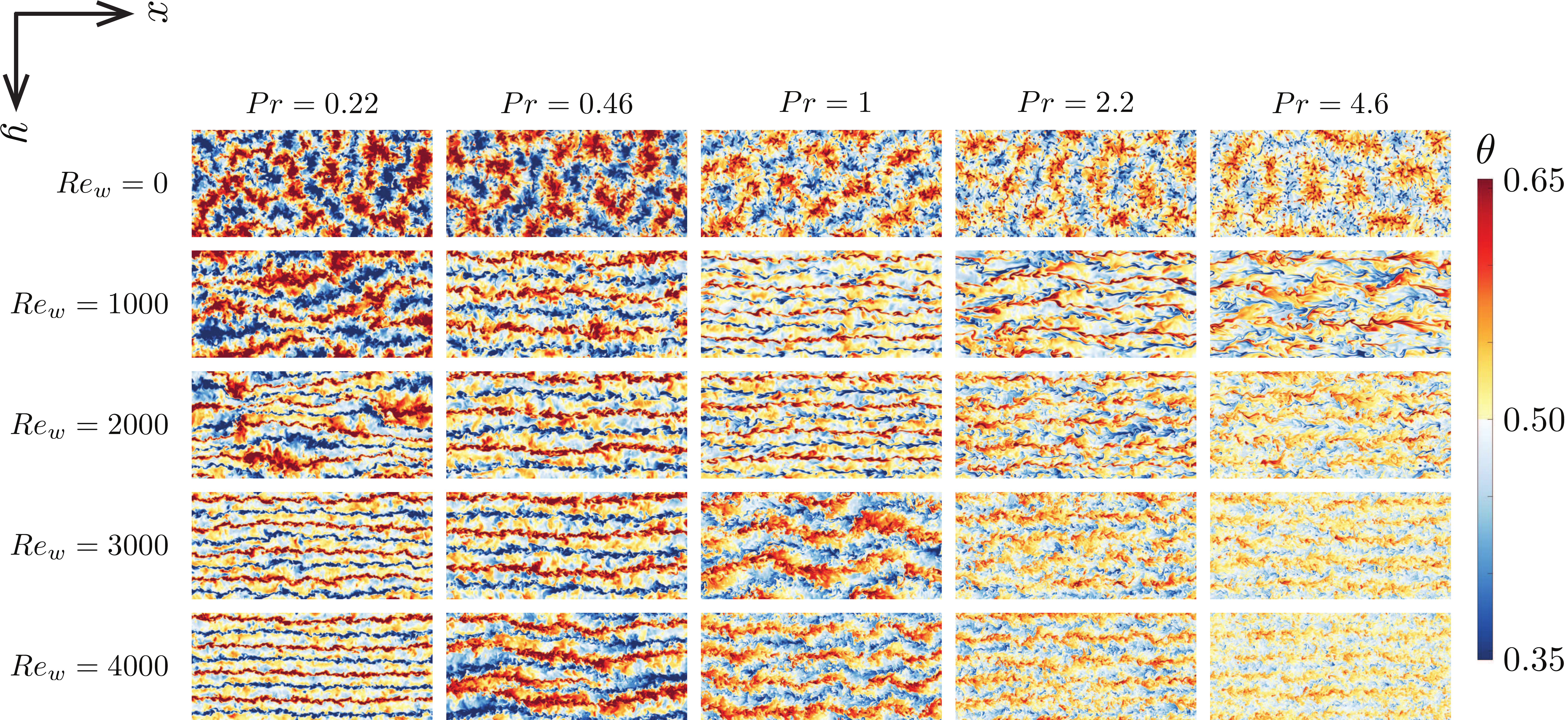}} 
	\caption{\label{fig:flowvisualizations} Snapshots of the temperature field at midheight ($z/H=0.5$) for a subdomain of the parameter space. The applied wall shear is in $x$-direction, while $y$ is the spanwise coordinate.}
\end{figure}

Following \cite{bla20}, we performed our simulations in a $9\pi H \times 4\pi H \times H$ domain, which are the streamwise, spanwise and wall-normal directions, respectively. The grid resolutions are also based on \cite{bla20} and then further modified to account for the Prandtl number variation in this study. \\ 

\section{Flow organization \& global transport properties}
\label{sec:results}
\subsection{Organization of turbulent structures}
\label{subsec:organization}

Using as guideline the description of \cite{bla20} we observe that also in the present case the flow can be classified in buoyancy dominated, transitional and shear dominated regimes (see figure \ref{fig:flowvisualizations} and Table \ref{tab:gridcases} for a full overview). As shown in \cite{bla20}, for $Pr=1$ and increasing $Re_w$, we observe the thermal buoyancy dominated regime at $Re_w =0$ while already at $Re_w =1000,2000$ the compact thermal structures elongate into streaks and evidence the transitional regime. Further increasing the wall shear causes the streaks to meander in the spanwise direction which indicates the shear dominated regime ($Re_w =3000,4000$). 

As $Pr=\nu/\kappa$ exceeds unity, kinematic viscosity overtakes thermal diffusivity and the wall shear affects the flow structures in the bulk earlier. In fact, it can be observed that already for $Re_w =1000$ the flow shows the meandering behaviour of the shear dominated regime. For $Pr =4.6$ and $Re_w =4000$ the shear is strong enough to make the effect of the thermal forcing negligible, as confirmed by the flow structures similar to the plane Couette flow. 

Conversely, for Prandtl numbers smaller than unity, the shear is less effective for a given $Re_w$ and the bulk flow is more dominated by the thermal structures. In the case of $Pr =0.22$, a wall shear of $Re_w =1000$ is not strong enough to fully disturb the plumes and only the next data point at $Re_w =2000$ shows signs of elongated streaks.

From the panels of figure \ref{fig:flowvisualizations} it is evident how $Pr$ changes the relative strength of momentum and thermal diffusivities: A higher Prandtl number increases the momentum transfer from the boundaries to the bulk and the transition to the shear dominated regime occurs at a lower $Re_w$ than for a corresponding low $Pr$ flow. Vice versa, for small Prandtl numbers, the thermal dominated regime is more persistent and the shear dominated flow features appear only at high $Re_w$.

\subsection{Heat transfer}
\label{subsec:heattransfer}

\begin{figure}
	\centering
	\includegraphics[width=\textwidth]{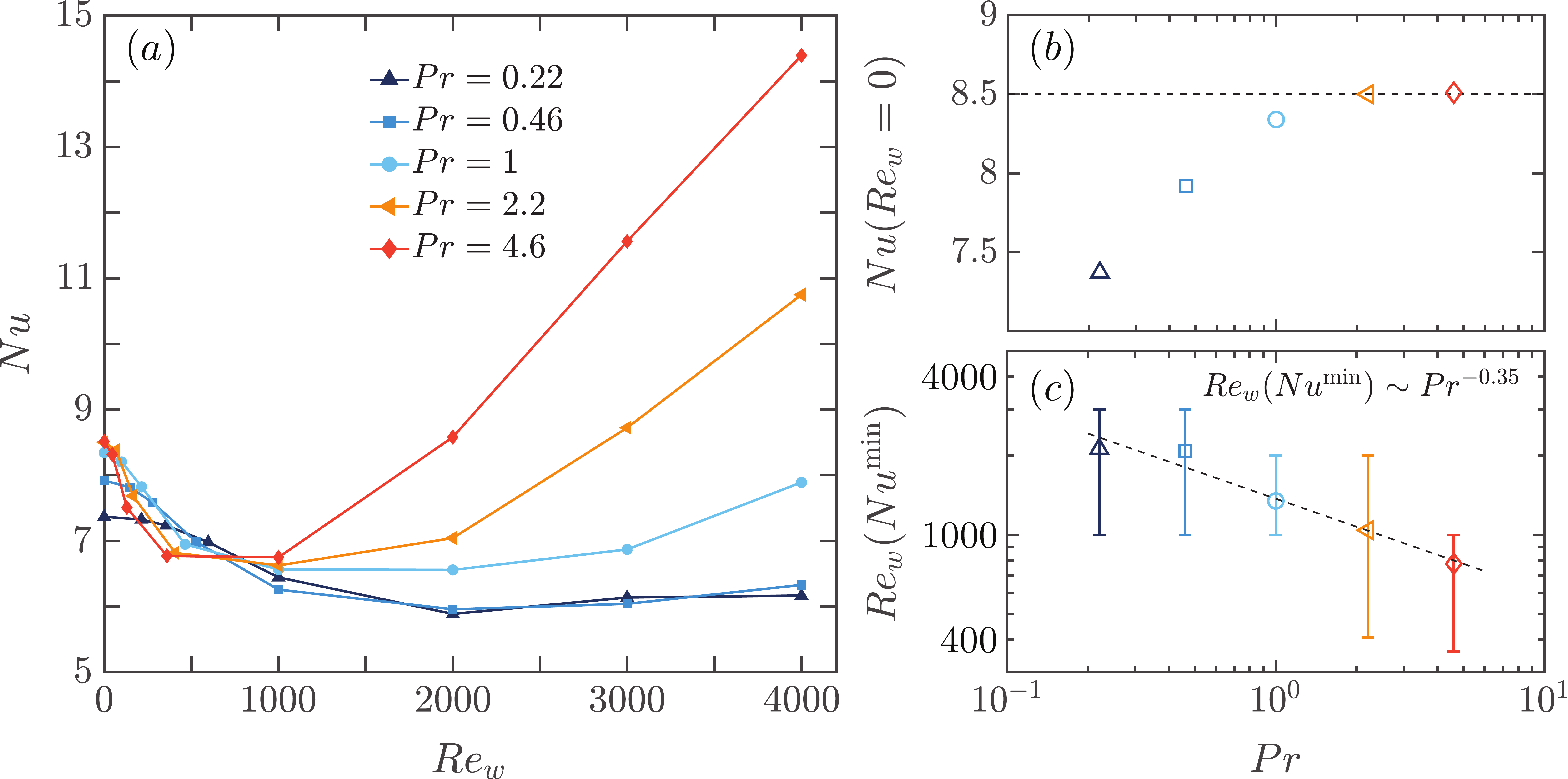}%
	\caption{\label{fig:NuRe} (a) $Nu$ versus $Re_w$ for varying $Pr$. The curves show a more or less pronounced minimum $Nu^{\min}$ at a certain shear Reynolds number $Re_w(Nu^{\min})$. (b) shows $Nu(Re_w=0)$ versus $Pr$. (c) shows $Re_w(Nu^{\min})$ versus $Pr$. Note that the error bars for these values are considerable, given our limited resolution in $Re_w$. Nonetheless, we include a power law fit into the figure.}
\end{figure}

The Nusselt number $Nu$ is plotted in figure \ref{fig:NuRe} as function of $Re_w$, showing a non-monotonic behaviour. The common feature is that for increasing wall shear, $Nu$ first decreases and then increases as already observed in \cite{bla20} for $Pr=1$. In the present case, however, the specific values are strongly dependent on $Pr$, as seen in figure \ref{fig:NuRe}c. The effect of $Pr$ is strongly dependent on the amount of shear added to the system. For pure Rayleigh-B\'enard convection ($Re_w=0$), $Nu$ increases with $Pr$ for $Pr<1$ and saturates to a constant value for $1<Pr<4.6$, see figure \ref{fig:NuRe}b, in agreement with the findings of \cite{poe13} and \cite{ste13}. For increasing $Re_w$, Prandtl number effects on the heat transfer are more pronounced, because of the higher momentum transfer from the boundaries to the bulk. This is confirmed both by the initial $Nu$ decrease up to $20\%$ of the RB value at $Pr=4.6$ and the subsequent strong increase by more than $50\%$ for the highest $Re_w$. In both cases the effects of the momentum transfer are enhanced by the high Prandtl number. We wish to stress that the non-monotonic behaviour of the Nusselt number observed here is a common feature of flows in which more than one parameter concur to determine the value of the heat transfer; for example similar dynamics are reported by \cite{yan20} and \cite{wan20} for thermal convection with rotation or \cite{cho16} for severe lateral confinement.

\subsection{Flow layering}
\label{subsec:layering}

\begin{figure}
	\centering 
	\includegraphics[width=\textwidth]{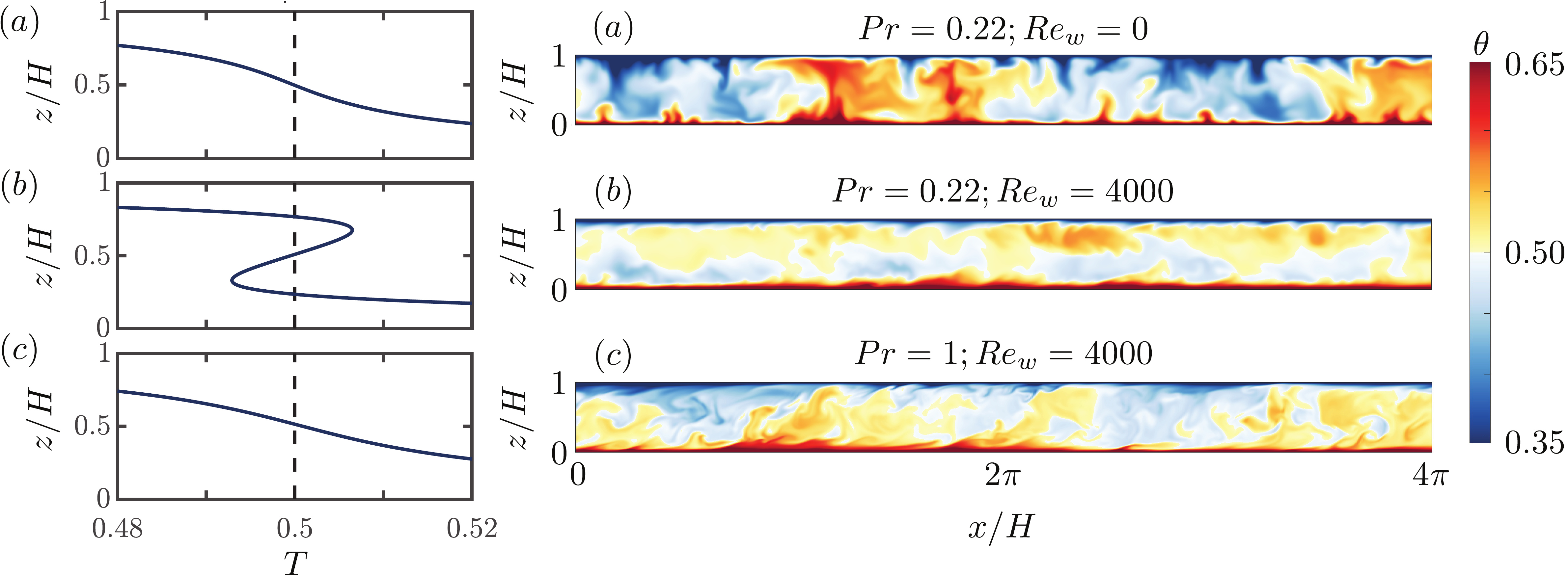}
	\caption{\label{fig:layering} Mean wall-normal temperature profiles (left) and side view snapshots of temperature fields (right), i.e. streamwise cross-sections, for (a) $Pr=0.22;Re_w=0$, (b) $Pr=0.22; Re_w=4000$, and (c) $Pr=1;Re=4000$. For all right panels only $x/H=0-4\pi$ is shown for better visibility and $y/H=2 \pi$ was chosen for the spanwise location, in which periodic boundary conditions are employed.}
\end{figure}

The initial $Nu$ decrease can be understood upon considering that the added wall shear perturbs the thermal RB structures and produces a horizontal flow layering that weakens the vertical heat flux. Once the wall shear is strong enough, however, the flow undergoes a transition to a shear dominated regime and the vertical cross-stream motion generated by the elongated streaks makes up for the suppressed RB structures, thus starting the Nusselt number monotonic increase \citep{bla20}. To better understand the effect of the horizontal flow layering, we discuss the results of figure \ref{fig:layering}. In these `side views' (i.e., streamwise cross-sections) of the temperature field snapshots and the corresponding top views of figure \ref{fig:flowvisualizations}, we can observe how the flow changes from thermal plumes to straight thin streaks and then to meandering structures. As expected, the increase in wall shear causes the flow to become more turbulent. But the change in the large-scale structures is also very recognizable. Here, the transitional regime displays a more unexpected behaviour. In contrast to what is seen in panels \ref{fig:layering}(a,c), where the flow structures appear clearly divided into hot and cold columns, in panel \ref{fig:layering}b the structures are more complex. Due to the wall shear and the thereby imposed horizontal flow, the vertical structures are disturbed, the flow is not able to reach the opposite hot/cold wall, but is instead trapped in a warm/cool state in the bulk of the flow. The fluctuations in the flow are not strong enough to mix the bulk and therefore the heat gets insulated in a stably stratified layer in the middle of the flow. This layering causes the total heat transfer to decrease and is the reason for the drop in $Nu$ for low $Re_w$ in figure \ref{fig:NuRe}. Because of the heat entrapment in the bulk layer, relatively cold fluid comes very close to relatively warm fluid and the temperature gradients in wall-normal direction increase significantly. In the atmosphere, this phenomenon can be observed as cloud streets, which, similar to the here observed high-shear end of the transitional regime, manifests as long streaks of convection rolls \citep{etl93,kim03,jay18}.

\section{Boundary layers}
\label{sec:BL}

\subsection{Boundary layer thicknesses}

\begin{figure}
	\centering 
	\includegraphics[width=\textwidth]{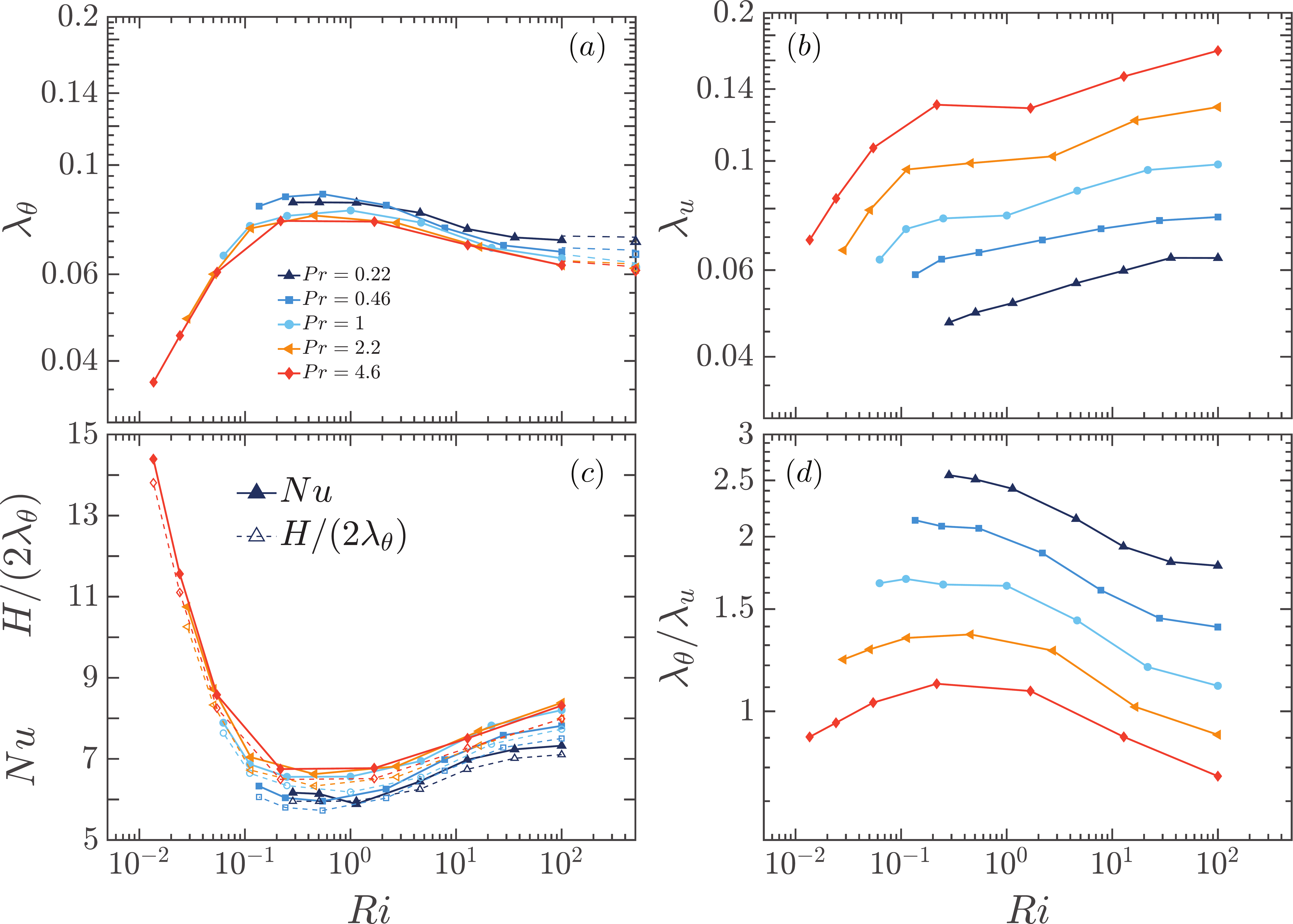}
	\caption{\label{fig:Ri_BL} (a) Thermal boundary layer thickness $\lambda_\theta$ and (b) kinetic boundary layer thickness $\lambda_u$ as function of the $Ri$-number for various $Pr$-numbers and fixed $Ra=10^6$. Note that the scale is the same in both (a) \& (b). (c) $Nu(Ri)$ compared to $H/(2\lambda_\theta (Ri))$. (d) Ratio of thermal and kinetic boundary layer thickness vs $Ri$.}
\end{figure}

A complementary way to better understand the $Pr$-dependence of the flow dynamics and the transport properties is to study the viscous and thermal boundary layer thicknesses $\lambda_u$ and $\lambda_\theta$, respectively. Here, we define both $\lambda_\theta$ and $\lambda_u$ by extrapolating the linear slopes of the mean temperature and mean streamwise velocity close to the walls, similarly to \cite{shi10}. The dependence of $\lambda_u$ and $\lambda_\theta$ on $Ri$ and $Pr$ is shown in figure \ref{fig:Ri_BL}. Here we use as abscissa the Richardson number. Given that $Ra=10^6$ is constant, we have $Ri \propto (PrRe_w^2)^{-1}$. At every $Pr$, for increasing $Ri$ -- and therefore decreasing shear -- $\lambda_\theta$ initially grows, then reaches a plateau around $Ri \approx 1$ and eventually decreases slowly to converge to the pure RB value (figure \ref{fig:Ri_BL}a). For comparison, we also plot $Nu(Ri)$ in figure \ref{fig:Ri_BL}c. Given that $\lambda_\theta \propto (Nu)^{-1}$ to a good approximation, the behaviour of the thermal boundary layer thickness is consistent with the Nusselt number of figures \ref{fig:NuRe} and \ref{fig:Ri_BL}c. The different flow regimes can be identified either from the different slopes of $\lambda_\theta$ versus $Ri$ or from those of $Nu(Ri)$. The slope is positive in the shear dominated region (small $Ri$), approximately zero in the transitional regime and then negative in the thermal buoyancy dominated regime.

As the Richardson number indicates the relative strength of buoyancy and shear, one might think that for increasing $Ri$ there should be a monotonic $\lambda_\theta$ decrease which, instead, is observed only for $Ri\gtrsim 1$. The reason for the counter-intuitive $\lambda_\theta$ increase for $Ri\lesssim 1$ is that in this region the thermal forcing is weak and the flow is mainly driven by the shear. In this case the thermal boundary layer is slaved to the viscous boundary layer which, according to the expectations, monotonically thickens as the wall shear weakens. From figure \ref{fig:Ri_BL}b we can see that indeed $\lambda _u$ monotonously increases with increasing $Ri$. 

Note that the viscous boundary layer thickness has a stronger dependence on $Pr$ than the thermal boundary layer thickness. Qualitatively, larger $Pr$ reflects stronger momentum diffusivity and therefore a thicker viscous boundary layer. In the shear dominated regime (high $Pr$ or low $Ri$), however, $\lambda_u$ grows faster than in the other regimes and this is especially true for the flows with higher $Pr$. In fact, in these cases the thermal boundary layer is nested within the viscous one and the dynamics of the latter is not influenced by the former. This is not the case for small $Pr < 1$ because then $\lambda_u$ evolves inside $\lambda_\theta$ whose thinning with increasing $Ri$ counteracts the thickening of the viscous boundary layer.

To further stress the importance of the relative thicknesses of the thermal and the viscous boundary layer, we show their ratio versus $Ri$ in figure \ref{fig:Ri_BL}d. We can see that $\lambda_\theta/\lambda_u$ increases for decreasing $Pr$ at fixed $Ri$ since the kinetic boundary layer thickness is driven by the momentum diffusivity. At fixed $Pr$ the behaviour of the boundary layer ratio is more complex: it always shows a decreasing trend in the high end of $Ri$ which is due to the thinning of the thermal boundary layer. On the other hand, at the low end of $Ri$ one can observe an increase only for $Pr > 1$, which is due to the steep growth of $\lambda_\theta$ with $Ri$ observed in figure \ref{fig:Ri_BL}a.

Due to the limited amount of datapoints, we cannot show a more detailed behaviour in the extreme case of pure shear forcing. In contrast, in the limit of pure Rayleigh-B\'enard convection we do observe the asymptotic trend for $\lambda_\theta/\lambda_u$; there the effect of the shear becomes very small (no imposed shear, all shear due to natural convection roll) and the ratio depends on $Pr$ only. This saturation occurs earlier for smaller $Pr$, because the thermal forcing dominates over the shear forcing at smaller $Ra$. 

\begin{figure}
	\centering
	\includegraphics[width=\textwidth]{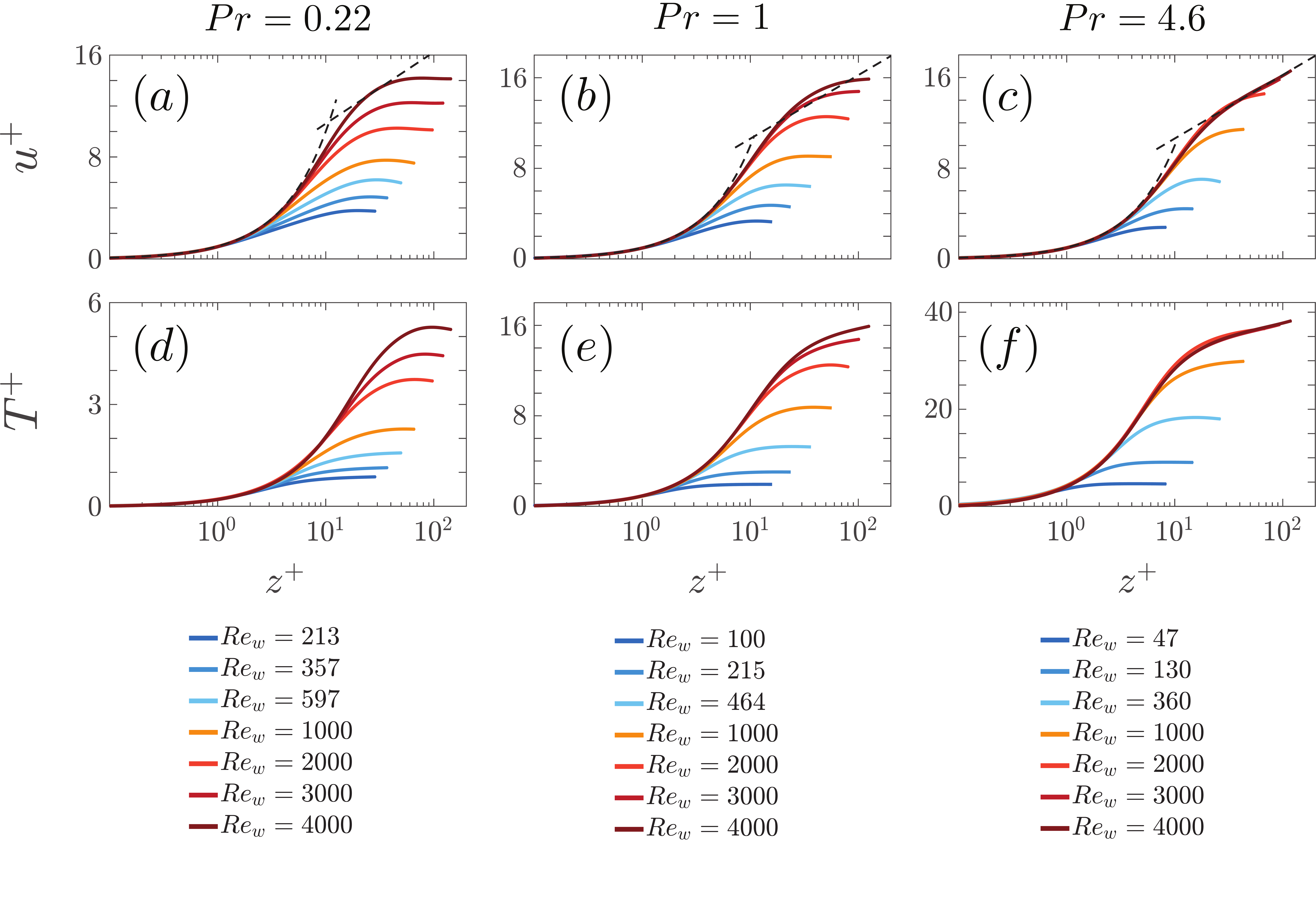}%
	\caption{\label{fig:flowstats} Velocity and temperature wall profiles for $Pr=0.22$ (left), $Pr=1$ (middle) and $Pr=4.6$ (right) for various $Re_w$. (a-c) Mean streamwise velocity and (d-f) mean temperature profiles. Here $u^+ =u/u_\tau$ and $T^+=T/T_\tau$, with the friction temperature $T_\tau =Q/u_\tau$. The dashed lines in (a-c) show the linear profile for $z^+<<10$ and the Prandtl- von Karman log-law of the wall $u^+(z^+)=\kappa ^{-1} \log z^+ +B$, with $\kappa= 0.41$ and $B=5$.}
\end{figure}

\subsection{Velocity and temperature wall profiles}
\label{subsec:veltemp}

For strong enough shear the boundary layers, which are first of laminar type, will eventually become turbulent, considerably enhancing the heat transport. However, for most of the values of the control parameters ($Re_w$ and $Pr$) of this paper this is hardly the case. This can best be judged from the velocity profiles, which we show in figures \ref{fig:flowstats}(a-c) for three different values of $Pr$ and various $Re_w$. Only in the high-$Pr$ range, towards the limit of plane Couette flow, we can see that $u^+$ evolves towards the well-known Prandtl- von Karman logarithmic behaviour $u^+(z^+)=\kappa^{-1} \log z^+ +B$ for high $Re_w$. Since the shear strongly affects the flow, the boundary layers can transition to turbulence earlier than without shear. But also the large $Pr$ enhance the shear. In fact, at $Pr=4.6$ already the flow at $Re_w=3000$ shows the onset of a log-law behaviour. The more $Pr$ is decreased, the harder it becomes for the wall shear to disturb the thermal plumes and, as a result, at $Re_w \leq 4000$ and $Pr \leq 2.2$, the log-scaling cannot be attained in our simulations. 

Panels \ref{fig:flowstats}(d-f) show a similar behaviour for the mean temperature profiles. One can observe that the temperature profiles converge earlier towards some type of logarithmic behaviour. For $Pr=1$, we can see such behaviour for $Re_w=4000$, whereas at larger $Pr=4.6$, it already shows up even at $Re_w=2000$. From the shown temperature profiles, we can also identify the flow layering that was previously discussed in section \ref{subsec:layering}. When the flow layering occurs, heat gets entrapped in the bulk flow. Since now an additional layer of warm and cool fluid exists in between of the cold and hot regions, $T^+$ shows a non-monotonic behaviour with a drop after the initial peak. This can most prominently be seen in figure \ref{fig:flowstats}d ($Pr=0.22$) for the strongest shear $Re_w=4000$.

\section{Conclusion}
\label{sec:conclusion}
In this manuscript we performed DNS of wall sheared thermal convection with $0\leq Re_w \leq 4000$ and $0.22 \leq Pr \leq 4.6$ at constant Rayleigh number $Ra=10^6$. Similarly to \cite{bla20}, who analysed the $Ra$-dependence of sall sheared thermal convection, we found three flow regimes and quantified them by using the bulk Richardson number and a visual analysis of two-dimensional cross-sectional snapshots. The flow undergoes a transition from the thermal buoyancy dominated to the transitional state when $Ri \lessapprox 10$. We found that the meandering streaks of the shear dominated regime start to emerge at $Ri \lessapprox 0.1$. Also the behaviour of the Nusselt number strongly depends on $Pr$. For high Prandtl number, the momentum transfer from the walls to the flow is increased and therefore the flow can easier reach the shear dominated regime where the heat transfer is again increased. We analysed both the thermal and the kinetic boundary layer thicknesses to better understand the transitions of the flow between its different regimes. We found that the thermal boundary layer thickness $\lambda_\theta$ shows a peak in the transitional regime and decreases for both lower and higher $Ri$. The kinetic boundary layer thickness $\lambda_u$ increases with increasing $Ri$ and increasing $Pr$. For very strong $Re_w$ and in particular large $Pr$ we notice the appearance of logarithmic boundary layer profiles, signalling the onset of turbulent boundary layer dynamics, leading to an enhanced heat transport.

Together with the results of \cite{bla20}, we now have analysed two orthogonal cross-sections of the three-dimensional parameter space $(Ra,Pr,Re_w)$. More specifically, we have determined $Nu(Ra,Pr,Re_w)$ for the two cross-sections $Nu(Ra,Pr=1,Re_w)$ in \cite{bla20} and $Nu(Ra=10^6,Pr,Re_w)$ here. From standard RB without shear we of course know $Nu(Ra,Pr,Re_w=0)$, which is perfectly described by the unifying theory of thermal convection by \cite{gro00,gro01} and \cite{ste13}. The knowledge of the two new cross-sections in parameter space may enable us to extend this unifying theory to sheared convection.

\section*{Acknowledgements}
\noindent We thank Pieter Berghout, Kai Leong Chong, and Olga Shishkina for fruitful discussions. The simulations were supported by a grant from the Swiss National Supercomputing Centre (CSCS) under project ID s713, s802, and s874. This work was financially supported by NWO and the Priority Programme SPP 1881 ``Turbulent Superstructures" of the Deutsche Forschungsgemeinschaft. We also acknowledge the Dutch national e-infrastructure SURFsara with the support of SURF cooperative.

\section*{Declaration of Interests}
\noindent The authors report no conflict of interest.

\bibliographystyle{jfm}
\bibliography{../../../../GitHub/Bibliography-turbulence/literature_turbulence}

\begin{thebibliography}{53}
\expandafter\ifx\csname natexlab\endcsname\relax\def\natexlab#1{#1}\fi
\def\au#1{#1} \def\ed#1{#1} \def\yr#1{#1}\def\at#1{#1}\def\jt#1{\textit{#1}}
  \def\bt#1{#1}\def\bvol#1{\textbf{#1}} \def\vol#1{#1} \def\pg#1{#1}
  \def\publ#1{#1}\def\arxiv#1{#1}\def\org#1{#1}\def\st#1{\textit{#1}}

\bibitem[Ahlers {\em et~al.\/}(2009)Ahlers, Grossmann \& Lohse]{ahl09}
{\sc \au{Ahlers, G.}, \au{Grossmann, S.} \& \au{Lohse, D.}} \yr{2009}  \at{Heat
  transfer and large scale dynamics in turbulent {{Rayleigh-B{\'e}nard}}
  convection}.  \jt{Rev. Mod. Phys.}  \bvol{81},  \pg{503--537}.

\bibitem[Barkley \& Tuckerman(2005)]{bar05}
{\sc \au{Barkley, D.} \& \au{Tuckerman, L.~S.}} \yr{2005}  \at{Computational
  study of turbulent laminar patterns in {Couette} flow}.  \jt{Phys. Rev.
  Lett.}  \bvol{94}~(1),  \pg{014502}.

\bibitem[Blass {\em et~al.\/}(2020)Blass, Zhu, Verzicco, Lohse \&
  Stevens]{bla20}
{\sc \au{Blass, A.}, \au{Zhu, X.}, \au{Verzicco, R.}, \au{Lohse, D.} \&
  \au{Stevens, R. J. A.~M.}} \yr{2020}  \at{Flow organization and heat transfer
  in turbulent wall sheared thermal convection}.  \jt{J. Fluid Mech.}
  \bvol{897},  \pg{A22}.

\bibitem[Chantry {\em et~al.\/}(2017)Chantry, Tuckerman \& Barkley]{cha17}
{\sc \au{Chantry, M.}, \au{Tuckerman, L.~S.} \& \au{Barkley, D.}} \yr{2017}
  \at{Universal continuous transition to turbulence in a planar shear flow}.
  \jt{J. Fluid Mech.}  \bvol{824},  \pg{R1}.

\bibitem[Chilla \& Schumacher(2012)]{chi12}
{\sc \au{Chilla, F.} \& \au{Schumacher, J.}} \yr{2012}  \at{New perspectives in
  turbulent {{Rayleigh-B{\'e}nard}} convection}.  \jt{Eur. Phys. J. E}
  \bvol{35},  \pg{58}.

\bibitem[Chong {\em et~al.\/}(2018)Chong, Wagner, Kaczorowski, Shishkina \&
  Xia]{cho18}
{\sc \au{Chong, K.~L.}, \au{Wagner, S.}, \au{Kaczorowski, M.}, \au{Shishkina,
  O.} \& \au{Xia, K.-Q.}} \yr{2018}  \at{Effect of prandtl number on heat
  transport enhancement in {Rayleigh-B\'enard} convection under geometrical
  confinement}.  \jt{Phys. Rev. Fluids}  \bvol{3},  \pg{013501}.

\bibitem[Chong \& Xia(2016)]{cho16}
{\sc \au{Chong, K.~L.} \& \au{Xia, K.-Q.}} \yr{2016}  \at{Exploring the
  severely confined regime in {Rayleigh-B\'enard} convection}.  \jt{J. Fluid
  Mech.}  \bvol{805},  \pg{R4}.

\bibitem[Deardorff(1972)]{dea72}
{\sc \au{Deardorff, J.~W.}} \yr{1972}  \at{Numerical investigation of neutral
  and unstable planetary boundary layers}.  \jt{J. Atmos. Sci.}  \bvol{29}~(1),
   \pg{91--115}.

\bibitem[Domaradzki \& Metcalfe(1988)]{dom88}
{\sc \au{Domaradzki, J.~A.} \& \au{Metcalfe, R.~W.}} \yr{1988}  \at{Direct
  numerical simulations of the effects of shear on turbulent
  {{Rayleigh-B{\'e}nard}} convection}.  \jt{J. Fluid Mech.}  \bvol{193},
  \pg{499}.

\bibitem[Etling \& Brown(1993)]{etl93}
{\sc \au{Etling, D.} \& \au{Brown, R.~A.}} \yr{1993}  \at{Roll vortices in the
  planetary boundary layer: A review}.  \jt{Bound.-Layer Meteorol.}  \bvol{65},
   \pg{215--248}.

\bibitem[Fukui \& Nakajima(1985)]{fuk85}
{\sc \au{Fukui, K.} \& \au{Nakajima, M.}} \yr{1985}  \at{Unstable
  stratification effects on turbulent shear flow in the wall region}.  \jt{Int.
  J. Heat Mass Transf.}  \bvol{28},  \pg{2343--2352}.

\bibitem[Grossmann \& Lohse(2000)]{gro00}
{\sc \au{Grossmann, S.} \& \au{Lohse, D.}} \yr{2000}  \at{Scaling in thermal
  convection: A unifying view}.  \jt{J. Fluid Mech.}  \bvol{407},  \pg{27--56}.

\bibitem[Grossmann \& Lohse(2001)]{gro01}
{\sc \au{Grossmann, S.} \& \au{Lohse, D.}} \yr{2001}  \at{Thermal convection
  for large {{Prandtl}} number}.  \jt{Phys. Rev. Lett.}  \bvol{86},
  \pg{3316--3319}.

\bibitem[Hathaway \& Somerville(1986)]{hat86}
{\sc \au{Hathaway, D.} \& \au{Somerville, R.}} \yr{1986}  \at{Nonlinear
  interactions between convection, rotation and flows with vertical shear}.
  \jt{J. Fluid Mech.}  \bvol{164},  \pg{91--105}.

\bibitem[Ingersoll(1966)]{ing66}
{\sc \au{Ingersoll, A.~P.}} \yr{1966}  \at{Thermal convection with shear at
  high {{Rayleigh}} number}.  \jt{J. Fluid Mech.}  \bvol{25},  \pg{209--228}.

\bibitem[Jayaraman \& Brasseur(2018)]{jay18}
{\sc \au{Jayaraman, B.} \& \au{Brasseur, J.~G.}} \yr{2018}  \at{The surprising
  transition in atmospheric boundary layer turbulence structure from neutral to
  moderately convective stability states and mechanisms underlying large-scale
  rolls}.  \jt{arXiv:1807.03336v2} .

\bibitem[Khanna \& Brasseur(1998)]{kha98}
{\sc \au{Khanna, S.} \& \au{Brasseur, J.~G.}} \yr{1998}  \at{Three-dimensional
  buoyancy- and shear-induced local structure of the atmospheric boundary
  layer}.  \jt{J. Atmos. Sci.}  \bvol{55}~(5),  \pg{710--743}.

\bibitem[Kim {\em et~al.\/}(2003)Kim, Park \& Moeng]{kim03}
{\sc \au{Kim, S.-W.}, \au{Park, S.-U.} \& \au{Moeng, C.-H.}} \yr{2003}
  \at{Entrainment processes in the convective boundary layer with varying wind
  shear}.  \jt{Bound.-Layer Meteorol.}  \bvol{108},  \pg{221--245}.

\bibitem[Kitoh \& Umeki(2008)]{kit08}
{\sc \au{Kitoh, O.} \& \au{Umeki, M.}} \yr{2008}  \at{Experimental study on
  large-scale streak structure in the core region of turbulent plane {Couette}
  flow}.  \jt{Phys. Fluids}  \bvol{20},  \pg{025107}.

\bibitem[Kooij {\em et~al.\/}(2018)Kooij, Botchev, Frederix, Geurts, Horn,
  Lohse, van~der Poel, Shishkina, Stevens \& Verzicco]{koo18}
{\sc \au{Kooij, G.~L.}, \au{Botchev, M.~A.}, \au{Frederix, E. M.~A.},
  \au{Geurts, B.~J.}, \au{Horn, S.}, \au{Lohse, D.}, \au{van~der Poel, E.~P.},
  \au{Shishkina, O.}, \au{Stevens, R. J .A.~M.} \& \au{Verzicco, R.}} \yr{2018}
   \at{Comparison of computational codes for direct numerical simulations of
  turbulent {{Rayleigh-B{\'e}nard}} convection}.  \jt{Computers $\&$ Fluids}
  \bvol{166},  \pg{1--8}.

\bibitem[Lee \& Moser(2018)]{lee18}
{\sc \au{Lee, M.} \& \au{Moser, R.~D.}} \yr{2018}  \at{Extreme-scale motions in
  turbulent plane {Couette} flows}.  \jt{J. Fluid Mech.}  \bvol{842},
  \pg{128--145}.

\bibitem[Lee \& Kim(1991)]{lee91}
{\sc \au{Lee, M.~J.} \& \au{Kim, J.}} \yr{1991}  \at{The structure of
  turbulence in a simulated plane {Couette} flow}.  \jt{Proceedings of the 8th
  Symposium on Turbulent Shear Flows, Munich}  \pg{pp. 5.3.1--5.3.6}.

\bibitem[Lohse \& Xia(2010)]{loh10}
{\sc \au{Lohse, D.} \& \au{Xia, K.-Q.}} \yr{2010}  \at{Small-scale properties
  of turbulent {{Rayleigh-B{\'e}nard}} convection}.  \jt{Annu. Rev. Fluid
  Mech.}  \bvol{42},  \pg{335--364}.

\bibitem[Moeng(1984)]{moe84}
{\sc \au{Moeng, C.-H.}} \yr{1984}  \at{A large-eddy-simulation model for the
  study of planetary boundary-layer turbulence}.  \jt{J. Atmos. Sci.}
  \bvol{41}~(13),  \pg{2052--2062}.

\bibitem[Orlandi {\em et~al.\/}(2015)Orlandi, Bernardini \& Pirozzoli]{orl15}
{\sc \au{Orlandi, P.}, \au{Bernardini, M.} \& \au{Pirozzoli, S.}} \yr{2015}
  \at{Poiseuille and {Couette} flows in the transitional and fully turbulent
  regime}.  \jt{J. Fluid Mech.}  \bvol{770},  \pg{424--441}.

\bibitem[Ostilla-M\'onico {\em et~al.\/}(2014)Ostilla-M\'onico, van~der Poel,
  Verzicco, Grossmann \& Lohse]{ost14d}
{\sc \au{Ostilla-M\'onico, R.}, \au{van~der Poel, E.~P.}, \au{Verzicco, R.},
  \au{Grossmann, S.} \& \au{Lohse, D.}} \yr{2014}  \at{Exploring the phase
  diagram of fully turbulent {Taylor-Couette} flow}.  \jt{J. Fluid Mech.}
  \bvol{761},  \pg{1--26}.

\bibitem[Pirozzoli {\em et~al.\/}(2011)Pirozzoli, Bernardini \& Orlandi]{pir11}
{\sc \au{Pirozzoli, S.}, \au{Bernardini, M.} \& \au{Orlandi, P.}} \yr{2011}
  \at{Large-scale motions and inner/outer layer interactions in turbulent
  {{Couette-Poiseuille}} flows}.  \jt{J. Fluid Mech.}  \bvol{680},
  \pg{534--563}.

\bibitem[Pirozzoli {\em et~al.\/}(2014)Pirozzoli, Bernardini \& Orlandi]{pir14}
{\sc \au{Pirozzoli, S.}, \au{Bernardini, M.} \& \au{Orlandi, P.}} \yr{2014}
  \at{Turbulence statistics in {Couette} flow at high {{Reynolds}} number}.
  \jt{J. Fluid Mech.}  \bvol{758},  \pg{327--343}.

\bibitem[Pirozzoli {\em et~al.\/}(2017)Pirozzoli, Bernardini, Verzicco \&
  Orlandi]{pir17}
{\sc \au{Pirozzoli, S.}, \au{Bernardini, M.}, \au{Verzicco, R.} \& \au{Orlandi,
  P.}} \yr{2017}  \at{Mixed convection in turbulent channels with unstable
  stratification}.  \jt{J. Fluid Mech.}  \bvol{821},  \pg{482--516}.

\bibitem[van~der Poel {\em et~al.\/}(2015)van~der Poel, Ostilla-M\'onico,
  Donners \& Verzicco]{poe15c}
{\sc \au{van~der Poel, E.~P.}, \au{Ostilla-M\'onico, R.}, \au{Donners, J.} \&
  \au{Verzicco, R.}} \yr{2015}  \at{A pencil distributed finite difference code
  for strongly turbulent wall-bounded flows}.  \jt{Computers $\&$ Fluids}
  \bvol{116},  \pg{10--16}.

\bibitem[van~der Poel {\em et~al.\/}(2013)van~der Poel, Stevens \&
  Lohse]{poe13}
{\sc \au{van~der Poel, E.~P.}, \au{Stevens, R. J. A.~M.} \& \au{Lohse, D.}}
  \yr{2013}  \at{Comparison between two and three dimensional
  {{Rayleigh-B{\'e}nard}} convection}.  \jt{J. Fluid Mech.}  \bvol{736},
  \pg{177--194}.

\bibitem[Scagliarini {\em et~al.\/}(2015)Scagliarini, Einarsson, Gylfason \&
  Toschi]{sca15}
{\sc \au{Scagliarini, A.}, \au{Einarsson, H.}, \au{Gylfason, A.} \& \au{Toschi,
  F.}} \yr{2015}  \at{Law of the wall in an unstably stratified turbulent
  channel flow}.  \jt{J. Fluid Mech.}  \bvol{781},  \pg{R5}.

\bibitem[Scagliarini {\em et~al.\/}(2014)Scagliarini, Gylfason \&
  Toschi]{sca14}
{\sc \au{Scagliarini, A.}, \au{Gylfason, A.} \& \au{Toschi, F.}} \yr{2014}
  \at{Heat-flux scaling in turbulent {{Rayleigh-B{\'e}nard}} convection with an
  imposed longitudinal wind}.  \jt{Phys. Rev. E}  \bvol{89},  \pg{043012}.

\bibitem[Shishkina {\em et~al.\/}(2010)Shishkina, Stevens, Grossmann \&
  Lohse]{shi10}
{\sc \au{Shishkina, O.}, \au{Stevens, R. J. A.~M.}, \au{Grossmann, S.} \&
  \au{Lohse, D.}} \yr{2010}  \at{Boundary layer structure in turbulent thermal
  convection and its consequences for the required numerical resolution}.
  \jt{New J. Phys.}  \bvol{12},  \pg{075022}.

\bibitem[Solomon \& Gollub(1990)]{sol90}
{\sc \au{Solomon, T.~H.} \& \au{Gollub, J.~P.}} \yr{1990}  \at{Sheared boundary
  layers in turbulent {{Rayleigh-B{\'e}nard}} convection}.  \jt{Phys. Rev.
  Lett.}  \bvol{64},  \pg{2382--2385}.

\bibitem[Stevens {\em et~al.\/}(2011)Stevens, Lohse \& Verzicco]{ste11}
{\sc \au{Stevens, R. J. A.~M.}, \au{Lohse, D.} \& \au{Verzicco, R.}} \yr{2011}
  \at{{{Prandtl}} and {{Rayleigh}} number dependence of heat transport in high
  {{Rayleigh}} number thermal convection}.  \jt{J. Fluid Mech.}  \bvol{688},
  \pg{31--43}.

\bibitem[Stevens {\em et~al.\/}(2013)Stevens, van~der Poel, Grossmann \&
  Lohse]{ste13}
{\sc \au{Stevens, R. J. A.~M.}, \au{van~der Poel, E.~P.}, \au{Grossmann, S.} \&
  \au{Lohse, D.}} \yr{2013}  \at{The unifying theory of scaling in thermal
  convection: {{The}} updated prefactors}.  \jt{J. Fluid Mech.}  \bvol{730},
  \pg{295--308}.

\bibitem[Stevens {\em et~al.\/}(2010)Stevens, Verzicco \& Lohse]{ste10}
{\sc \au{Stevens, R. J. A.~M.}, \au{Verzicco, R.} \& \au{Lohse, D.}} \yr{2010}
  \at{Radial boundary layer structure and {{Nusselt}} number in
  {{Rayleigh-B{\'e}nard}} convection}.  \jt{J. Fluid Mech.}  \bvol{643},
  \pg{495--507}.

\bibitem[Teimurazov \& Frick(2017)]{tei17}
{\sc \au{Teimurazov, A.} \& \au{Frick, P.}} \yr{2017}  \at{Thermal convection
  of liquid metal in a long inclined cylinder}.  \jt{Phys. Rev. Fluids}
  \bvol{2},  \pg{113501}.

\bibitem[Thurlow \& Klewicki(2000)]{thu00}
{\sc \au{Thurlow, E.~M.} \& \au{Klewicki, J.~C.}} \yr{2000}  \at{Experimental
  study of turbulent {{Poiseuille-Couette}} flow}.  \jt{Phys. Fluids}
  \bvol{12},  \pg{865--875}.

\bibitem[Tsukahara {\em et~al.\/}(2006)Tsukahara, Kawamura \& Shingai]{tsu06}
{\sc \au{Tsukahara, T.}, \au{Kawamura, H.} \& \au{Shingai, K.}} \yr{2006}
  \at{{{DNS}} of turbulent {Couette} flow with emphasis on the large-scale
  structure in the core region}.  \jt{J. Turb.}  \bvol{7},  \pg{N19}.

\bibitem[Tuckerman \& Barkley(2011)]{tuc11}
{\sc \au{Tuckerman, L.~S.} \& \au{Barkley, D.}} \yr{2011}  \at{Patterns and
  dynamics in transitional plane {Couette} flow}.  \jt{Phys. Fluids}
  \bvol{23}~(4),  \pg{041301}.

\bibitem[Usanov {\em et~al.\/}(1999)Usanov, Pankratov, Popov, Markelov, Ryabaya
  \& Zabrodskaya]{usa99}
{\sc \au{Usanov, V.~I.}, \au{Pankratov, D.~V.}, \au{Popov, {{\'e}}.~P.},
  \au{Markelov, P.~I.}, \au{Ryabaya, L.~D.} \& \au{Zabrodskaya, S.~V.}}
  \yr{1999}  \at{Long-lived radionuclides of sodium, lead-bismuth, and lead
  coolants in fast-neutron reactors}.  \jt{Atomic Energy}  \bvol{87},
  \pg{658--662}.

\bibitem[Verzicco \& Camussi(1997)]{ver97}
{\sc \au{Verzicco, R.} \& \au{Camussi, R.}} \yr{1997}  \at{Transitional regimes
  of low-{{Prandtl}} thermal convection in a cylindrical cell}.  \jt{Phys.
  Fluids}  \bvol{9},  \pg{1287--1295}.

\bibitem[Verzicco \& Camussi(2003)]{ver03}
{\sc \au{Verzicco, R.} \& \au{Camussi, R.}} \yr{2003}  \at{Numerical
  experiments on strongly turbulent thermal convection in a slender cylindrical
  cell}.  \jt{J. Fluid Mech.}  \bvol{477},  \pg{19--49}.

\bibitem[Verzicco \& Orlandi(1996)]{ver96}
{\sc \au{Verzicco, R.} \& \au{Orlandi, P.}} \yr{1996}  \at{A finite-difference
  scheme for three-dimensional incompressible flow in cylindrical coordinates}.
   \jt{J. Comput. Phys.}  \bvol{123},  \pg{402--413}.

\bibitem[Vignarooban {\em et~al.\/}(2015)Vignarooban, Xu, Arvay, Hsu \&
  Kannan]{vig15}
{\sc \au{Vignarooban, K.}, \au{Xu, Xinhai}, \au{Arvay, A.}, \au{Hsu, K.} \&
  \au{Kannan, A.M.}} \yr{2015}  \at{Heat transfer fluids for concentrating
  solar power systems – a review}.  \jt{Appl. Energy}  \bvol{146},
  \pg{383--396}.

\bibitem[Wang {\em et~al.\/}(2020)Wang, Chong, Stevens, Verzicco \&
  Lohse]{wan20}
{\sc \au{Wang, Q.}, \au{Chong, K.~L.}, \au{Stevens, R. J. A.~M.}, \au{Verzicco,
  R.} \& \au{Lohse, D.}} \yr{2020}  \at{From zonal flow to convection rolls in
  {Rayleigh-B\'enard} convection with free-slip plates}.  \jt{arXiv:2005.02084}
  .

\bibitem[Xia(2013)]{xia13}
{\sc \au{Xia, K.-Q.}} \yr{2013}  \at{Current trends and future directions in
  turbulent thermal convection}.  \jt{Theor. Appl. Mech. Lett.}  \bvol{3},
  \pg{052001}.

\bibitem[Yang {\em et~al.\/}(2020)Yang, Verzicco, Lohse \& Stevens]{yan20}
{\sc \au{Yang, Y.}, \au{Verzicco, R.}, \au{Lohse, D.} \& \au{Stevens, R. J.
  A.~M.}} \yr{2020}  \at{What rotation rate maximizes heat transport in
  rotating {{Rayleigh-B{\'e}nard}} convection with prandtl number larger than
  one?}  \jt{Phys. Rev. Fluids}  \bvol{5},  \pg{053501}.

\bibitem[Zhou {\em et~al.\/}(2017)Zhou, Taylor \& Caulfield]{zho17}
{\sc \au{Zhou, Q.}, \au{Taylor, J.~R.} \& \au{Caulfield, C.P.}} \yr{2017}
  \at{Self-similar mixing in stratified plane {Couette} flow for varying
  {Prandtl} number}.  \jt{J. Fluid Mech.}  \bvol{820},  \pg{86--120}.

\bibitem[Zhu {\em et~al.\/}(2018)Zhu, Phillips, Arza, Donners, Ruetsch, Romero,
  Ostilla-M\'onico, Yang, Lohse, Verzicco, Fatica \& Stevens]{zhu18b}
{\sc \au{Zhu, X.}, \au{Phillips, E.}, \au{Arza, V.~S.}, \au{Donners, J.},
  \au{Ruetsch, G.}, \au{Romero, J.}, \au{Ostilla-M\'onico, R.}, \au{Yang, Y.},
  \au{Lohse, D.}, \au{Verzicco, R.}, \au{Fatica, M.} \& \au{Stevens, R. J.
  A.~M.}} \yr{2018}  \at{{{AFiD-GPU}}: a versatile {{Navier-Stokes}} solver for
  wall-bounded turbulent flows on {{GPU}} clusters}.  \jt{Comput. Phys.
  Commun.}  \bvol{229},  \pg{199--210}.

\bibitem[Zonta \& Soldati(2018)]{zon18}
{\sc \au{Zonta, F.} \& \au{Soldati, A.}} \yr{2018}  \at{Stably stratified
  wall-bounded turbulence}.  \jt{Appl. Mech. Rev.}  \bvol{70},  \pg{040801}.

\end{thebibliography}

\end{document}